\documentstyle[11pt,psfig]{article}

\title{The Ab Initio Melting Curve of Aluminium} 

\author{ Lidunka Vo\v{c}adlo$^1$ and Dario Alf\`e$^{1,2}$\\
\small{$^1$Dept. of Geological Sciences, University College London,
Gower Street, London, WC1E 6BT}\\ \small{$^2$Dept. of Physics and
Astronomy,University College London, Gower Street, London, WC1E 6BT} }

\begin{document}
\maketitle

\begin{abstract}
The melting curve of aluminium has been determined from 0 to $\sim
$150 GPa using first principles calculations of the free energies of
both the solid and liquid. The calculations are based on density
functional theory within the generalised gradient approximation using
ultrasoft Vanderbilt pseudopotentials. The free energy of the harmonic
solid has been calculated within the quasiharmonic approximation using
the small-displacement method; the free energy of the liquid and the
anharmonic correction to the free energy of the solid have been
calculated {\em via} thermodynamic integration from suitable reference
systems, with thermal averages calculated using {\em ab-initio}
molecular dynamics.  The resulting melting curve is in good agreement
with both static compression measurements and shock data.

\end{abstract}

\section{Introduction}

The determination of the melting curves of materials to very high pressures is of fundamental importance to our understanding
of the properties of planetary interiors; however, obtaining such melting curves remains
a major challenge to experimentalists and theorists alike. In
particular, the melting behaviour of iron is of great interest to the
Earth science community, since knowledge of this melt transition would
help constrain the temperature at the inner core boundary (about 1200
km from the centre of the Earth) which is currently uncertain to
within a few thousand degrees. Although several attempts have been
made to obtain the melting curve of iron, experimentally and
theoretically determined melting curves vary widely with significant
disagreement between static compression
measurements~\cite{williams87,boehler93,shen98}, shock
data~\cite{brown86,yoo93} and first principles
calculations~\cite{alfe99,alfe01a,alfe01b,laio00,belonoshko00}. Consequently,
the true nature of the melting curve of iron remains in some dispute.

In order to test the reliability of the theoretical techniques used in
our previous work on iron and to validate further the reported melting
curve~\cite{alfe99,alfe01b}, we have calculated the melting curve of
aluminium, for which there is a plethora of ambient experimental data
(e.g. ~\cite{crc97}), and for which the experimental melting curve has
recently been measured~\cite{boehler97,hanstrom00,shaner84}.

In the past, a number of theoretical approaches have been used to
investigate the melting behaviour of aluminium.  Moriarty {\em et
al.}~\cite{moriarty84} used the generalised pseudopotential theory
(GPT) to calculate the free energy of both the solid and liquid. They
treated the solid harmonically within the quasiharmonic approximation
and for the liquid they used fluid variational theory, where an upper
bound for the free energy is calculated from a reference system
constructed within GPT. They obtained a melting curve to 200 GPa in
fair agreement with more recently determined experiment
data~\cite{boehler97,hanstrom00,shaner84}, predicting a zero pressure
melting temperature of 1050K compared to the experimental value of
933K~\cite{crc97}.
Mei and Davenport~\cite{mei92} used the embedded atom model (EAM)
based on an analytical potential fitted to the structural properties
of aluminium. They calculated the free energies of the solid and
liquid and obtained a melting temperature at zero pressure of
800K. Morris {\em et al.}~\cite{morris94} employed the same EAM model
but they used phase coexistence to determine the melting temperature
as a function of pressure, with results considerably lower than
previous theoretical and experimental estimates; they obtained a zero
pressure melting temperature of $\sim $720 K. Straub
{\em et al.}~\cite{straub94} used first principles calculations to construct
an optimal classical potential, and used this potential to calculate
the free energies of the solid and the liquid using molecular
dynamics; they obtained a zero pressure melting temperature of 955 K.

The first fully {\em ab-initio} determination of aluminium melting
behaviour is that of de Wijs {\em et al.}~\cite{dewijs98}, who
obtained the zero pressure melting point by calculating the free
energy of the solid and the liquid entirely from first
principles. Their calculations were based on density functional theory
(DFT)~\cite{generaldft} using the local-density approximation (LDA)
for the exchange-correlation energy.  The free energy of the solid was
obtained as the sum of the free energy of the harmonic solid, within
the quasiharmonic approximation, and the full anharmonic contribution,
calculated using thermodynamic integration~\cite{frenkel96} using the
harmonic solid as the reference system. For the liquid they used
thermodynamic integration with a Lennard-Jones fluid as the reference
system. They obtained a melting temperature of 890 K.  More recently,
Jesson and Madden~\cite{jesson2000} used the orbital-free (OF) variant
of {\em ab-initio} molecular dynamics and thermodynamic integration to
calculate the free energy of liquid and solid aluminium. They found a
melting temperature of 615 K, attributing the discrepancy with the
DFT-LDA value of de Wijs {\em et al.}~\cite{dewijs98} to either the OF
approximation or the pseudopotential used.

In this paper we present the first fully {\em ab-initio} calculations
of the entire melting curve of aluminium from 0 to 150 GPa. Our
calculations are similar in the general principles to those of de Wijs
{\em et al.}~\cite{dewijs98} in the sense that we calculate the {\em
ab-initio} free energies of both liquid and solid using thermodynamic
integration, although we use the generalised gradient approximation
(GGA)~\cite{wang91,perdew92} for the exchange-correlation energy. In
addition to extending the calculations to a wide range of pressures,
we also present a more efficient approach to the thermodynamic
integration scheme, in which additional intermediate steps are
introduced in order to minimise the computational effort. Finally, we
discuss some possible limitations of the GGA.

The paper is organised as follows: in section~\ref{sec:techniques} we
describe the {\em ab-initio} simulation techniques and the strategy to
calculate the melting curve; in section~\ref{sec:liquid}
and~\ref{sec:solid} we describe the calculations of the free energy of
the liquid and the solid respectively, and in
section~\ref{sec:discussion} we present the melting properties of
aluminium.

\section{{\em Ab-initio} simulation techniques and strategy for melting}\label{sec:techniques}

In the present work, the aluminium system was represented by a
collection of Al$^{3+}$ ions and $3N$ electrons, where $N$ is the
number of atoms. The ions were treated as classical particles, and
their motion was adiabatically decoupled from that of the electrons
{\em via} the Born-Oppenheimer approximation. For each position of the
ions, the electronic problem was solved within the framework of
DFT~\cite{generaldft} using the GGA of Perdew and Wang
~\cite{wang91,perdew92}. Thermal electronic excitations were included
using the standard methods of finite-temperature DFT developed by
Mermin~\cite{mermin65,gillan89,wentzcovitch92}.
The present calculations were performed with the code
VASP~\cite{kresse96a} which is exceptionally efficient for metals.
The interaction between electrons and nuclei was described with the
ultrasoft pseudopotential (USPP) method~\cite{vanderbilt90}. We used
plane-waves with a cut-off of 130~eV.  The Brillouin-zone was sampled
using Monkhorst-Pack special points~\cite{monkhorst76} (the detailed
form of sampling will be noted where appropriate). The extrapolation
of the charge density from one step to the next in the {\em ab-initio}
molecular dynamics (AIMD) simulations was performed using the
technique described by Alf\`e~\cite{alfe99b}, which improves the
efficiency of the calculations by almost a factor of two. The time
step used in our simulations was 1 femto-second.

To calculate the melting temperature we calculated the Gibbs free
energy of both the solid and the liquid as a function of pressure and
temperature, $G_s(P,T)$ and $G_l(P,T)$ and, at each chosen $P$,
obtained the melting temperature, $T_m$, from $G_s(P,T_m) = G_l(P,T_m)$.
In fact, we calculated the Helmholtz free energy $F(V,T)$ as a
function of volume and temperature, and the Gibbs free energy was
obtained from the usual expression $ G = F + PV $, where $P =
-(\partial F / \partial V)_T$ is the pressure.  The main problem in
determining melting curves with this technique is the high precision
with which the free energies need to be calculated. This is because
the Gibbs free energy of the liquid crosses the Gibbs free energy of the
solid at a shallow angle, the difference in the slopes being the
entropy change on melting. For aluminium this is about 1.4 $k_{\rm
B}$/atom at zero pressure, which means that an error of 0.01 eV/atom
in either $G_s$ or $G_l$ results in an error of $\approx 80$ K in the
melting temperature. Therefore, it is important to reduce
non-cancelling errors between the liquid and the solid to an absolute
minimum. In the next sections we give a detailed discussion of the
techniques that we have used to calculate the free energies of the
liquid and the solid, and report what the {\em controllable} errors are:
those due to {\bf k}-point sampling, finite size, and statistical
sampling. We also try to give an estimate of what the {\em
uncontrollable} errors due to DFT-GGA may be.


\section{Free energy of the liquid}\label{sec:liquid}

The Helmholtz free energy $F$ of a classical system containing $N$
particles is:
\begin{equation}
\label{eqn:helmholtz}
F = - k_{\rm B} T  \ln \left\{ \frac{1}{N ! \Lambda^{3 N}}
\int_V d {\bf R}_1 \ldots d {\bf R}_N \,
e^{ - \beta U ( {\bf R}_1 , \ldots {\bf R}_N ; T )}
 \right\},
\end{equation}
where $\Lambda = h / ( 2
\pi M k_{\rm B} T )^{1/2}$ is the thermal wavelength, with $M$ the
nuclear mass, $h$ the Plank's constant, $k_{\rm B}$ the Boltzmann constant
and $\beta = 1 / k_{\rm B} T$. The multidimensional integral extends
over the total volume of the system $V$.

A direct calculation of $F$ using the equation above is impossible,
since it would involve knowledge of the potential energy $ U (
{\bf R}_1 , \ldots {\bf R}_N ; T )$ for all possible positions of the
$N$ atoms in the system. We have used instead the technique known as
thermodynamic integration~\cite{frenkel96}, as developed in earlier
papers~\cite{sugino95,smargiassi95a,dewijs98,alfe00}.  This is a
general scheme to compute the free energy difference $F - F_0$ between
two systems whose potential energies are $U$ and $U_0$ respectively.
In what follows we will assume that $F$ is the unknown free energy of
the {\em ab-initio} system and $F_0$ is the known free energy of a
reference system.  The free energy difference $F - F_0$ is the
reversible work done when the potential energy function $U_0$ is
continuously and reversibly switched to $U$.  To do this switching, a
continuously variable energy function $U_\lambda$ is defined such that
for $\lambda=0, U_\lambda=U_0$ and for $\lambda=1, U_\lambda=U$. We
also require $U_\lambda$ to be differentiable with respect to
$\lambda$ for $0 \le \lambda \le 1$. A convenient form is:
\begin{equation}
U_\lambda = ( 1 - f(\lambda) ) U_0 + f(\lambda) U,
\end{equation}
where $f(\lambda)$ is an arbitrary continuous and differentiable
function of $\lambda$ with the property $f(0)=0$ and $f(1)=1$. The
Helmholtz free energy of this {\em hybrid} system is:
\begin{equation}
F_\lambda = - k_{\rm B} T  \ln \left\{ \frac{1}{N ! \Lambda^{3 N}}
\int_V d {\bf R}_1 \ldots d {\bf R}_N \,
e^{- \beta U_\lambda ( {\bf R}_1 , \ldots {\bf R}_N ; T )}
\right\},
\end{equation}
Differentiating this with respect to $\lambda $ gives:
\begin{equation}
\frac{dF_\lambda }{d\lambda }=-k_BT\frac{\frac {1}{N!\Lambda^{3N}}\int_V
d{\bf R}_1 \ldots d{\bf R}_N e^{-\beta U_\lambda ({\bf R}_1, \ldots
{\bf R}_N;T)}(-\beta\frac{\partial U_\lambda }{\partial \lambda })}{{\frac
{1}{N!\Lambda^{3N}}\int_V d{\bf R}_1 \ldots d{\bf R}_N e^{-\beta U_\lambda
({\bf R}_1, \ldots {\bf R}_N;T)}}}=\left\langle \frac{\partial U_\lambda
}{\partial \lambda } \right\rangle_\lambda,
\end{equation}
so
\begin{equation}
\label{eqn:ti}
\Delta F=F-F_0=\int\limits_0^1d\lambda \left\langle \frac{\partial
U_\lambda }{\partial \lambda }\right\rangle _\lambda .
\end{equation}
For our calculations we defined $U_\lambda $ thus:
\begin{equation}
U_\lambda =\left( 1-\lambda \right) U_0 + \lambda U.
\end{equation}
Differentiating U$_\lambda $ with respect to $\lambda $ and substituting
into Equation~\ref{eqn:ti} yields:
\begin{equation}\label{eqn:ti_with_linear_switching}
\Delta F=\int\limits_0^1d\lambda \left\langle U-U_0\right\rangle
_\lambda.
\end{equation}
Under the ergodicity hypothesis, thermal averages are equivalent to time
averages, so we calculated $ \left\langle \cdot \right\rangle _\lambda
$ using AIMD, taking averages over time, with the evolution of the system
determined by the potential energy function $U_\lambda$. The
temperature was controlled using a Nos\'e
thermostat~\cite{nose84,ditolla93}.  It is important to stress that
the choice of the reference system does not affect the final answer
for $F$, although it does affect the efficiency of the calculations.
The latter can be understood by analysing the quantity $\langle U -
U_0 \rangle_\lambda$.  If this difference has large fluctuations then
one would need very long simulations to calculate the average value to
a sufficient statistical accuracy. Moreover, for an unwise choice of
$U_0$ the quantity $\langle U - U_0 \rangle_\lambda$ may strongly
depend on $\lambda$ so that one would need a large number of
calculations at different $\lambda$'s in order to compute the integral
in Eq.~\ref{eqn:ti_with_linear_switching} with sufficient accuracy. It
is crucial, therefore, to find a good reference system, where "good"
means a system for which the fluctuations of $U - U_0$ are as small as
possible.  In fact, if the fluctuations are small enough, we can
simply write $F - F_0 \simeq \langle U - U_0 \rangle_0$, with the
average taken in the reference ensemble. If this is not good enough,
the next approximation is readily shown to be~\cite{alfe01a}:
\begin{equation}
F - F_0 \simeq \langle U -
U_0 \rangle_0 - \frac{1}{2 k_{\rm B} T}
\left\langle \left[ U - U_0 -
\langle U - U_0 \rangle_0 
\right]^2 \right\rangle_0.
\label{eqn:secondorder}
\end{equation}
This form is particularly convenient since one only needs to sample
the phase space with the reference system, and perform a number of
{\em ab-initio} calculations on statistically independent configurations
extracted from a long classical simulation.

To evaluate the integral in
Equation~\ref{eqn:ti_with_linear_switching} one can calculate the
integrand $\langle U - U_0 \rangle_\lambda$ at a sufficient number
of $\lambda$ and calculate the integral numerically.

Alternatively, one can adopt the dynamical method described by
Watanabe and Reinhardt~\cite{watanabe90}. In this approach the
parameter $\lambda$ depends on time, and is slowly (adiabatically)
switched from 0 to 1 during a single simulation.  The switching rate
has to be slow enough so that the system remains in thermodynamic
equilibrium, and adiabatically transforms from the reference to the
{\em ab-initio} system.  The change in free energy is then given by:
\begin{equation}\label{eqn:adiabatic_switching}
\Delta F=\int\limits_0^{T_{\rm sim}} dt\frac{d\lambda }{dt}\left(
U-U_0\right),
\end{equation}
where $T_{\rm sim}$ is the total simulation time, $\lambda(t)$ is an
arbitrary function of $t$ with the property of being continuous and
differentiable for $0\le t \le 1$,  $\lambda(0)=0$ and
$\lambda(T_{\rm sim}) = 1$.

When using this second method, it is important to ensure that the switching is
adiabatic, i.e. that $T_{\rm sim}$ is sufficiently large. This can be achieved by changing
$\lambda$ from 0 to 1 in the first half of the simulation, and then
from 1 back to 0 in the second half of the simulation, evaluating
$\Delta F$ in each case; the average of the two values is then taken as
the best estimate for $ \Delta F$, and the difference is a measure of the non-adiabaticty. If this difference is less than the desired statistical uncertainty, one can be confident that the simulation time is sufficiently long. 

In our calculations we chose a total simulation time of
sufficient length such that the difference in $\Delta F$
between the two calculations was less than a few meV/atom. We return later to estimate the
errors in our calculations in Section~\ref{sec:liquid_errors}.

As pointed out by Jesson and Madden~\cite{jesson2000}, a possible
problem in the calculation of the thermodynamic integral is that the
system $U_\lambda$ may be in the solid region of the phase diagram,
even though the two end members $U_0$ and $U$ are in the liquid
region.  If this happens, the system can freeze during the
switching, and the integration path is not reversible, leading to an
incorrect result.  For small systems the situation is even more
problematic, since the phase diagram is not defined by sharp
boundaries, and the system can freeze even if it is above the melting
temperature of the corresponding system in the thermodynamic limit.
We have ourselves experienced freezing of the system for some
simulations at temperatures very close to the melting point; in order to avoid
including the results from these simulations, we carefully
monitored the mean square displacement and the structure factor of the
system, and included only those simulations in which these two quantities
clearly indicated liquid behaviour throughout the whole simulation.

\subsection{The reference system}

We mentioned earlier that the efficiency of the calculations is
entirely determined by the quality of the reference system, i.e. by
the strength of the fluctuations of $\Delta U = U - U_0$.  The key to
the success of these simulations, therefore, is being able to find a
reference system such that the fluctuations in $\Delta U$ are as small
as possible.  Based on the experience of previous work on liquid
Al~\cite{dewijs98} and liquid Fe~\cite{alfe99,alfe01b} we experimented
with the Lennard-Jones (LJ) system and an inverse power potential
(IP). Analysis of the fluctuations in $\Delta U$ indicated that the
system which best represented the liquid was the IP:
\begin{equation}
U_{\rm IP} = \frac{1}{2} \sum_{I \ne J} \phi ( \mid {\bf R}_I -
{\bf R}_J \mid ),
\label{eqn:uip}
\end{equation}
where 
\begin{equation}
\phi(r)= \frac {B} {r^\alpha}. 
\end{equation}
The potential parameters $B$ and $\alpha$ were chosen by minimising
the quantity:
\begin{equation} \label{eqn:fluctuations}
\left\langle \left [ U_{\rm IP}\left( B ,\alpha\right) - U - \langle
U_{\rm IP}\left( B ,\alpha\right) - U \rangle \right ]^2\right\rangle
\end{equation}
with respect to $B$ and $\alpha$, where $\langle \rangle$ means
the thermal average in the ensemble generated by the {\em ab-initio}
potential.  To investigate whether the optimum values for the
potential parameters depended strongly on thermodynamic state, we
performed the optimisation at the three thermodynamic states of the
extremes of high $P/T$ and low $P/T$, and also a point in between; we
found that the single choice of $B =246.67$ and $\alpha=6.7$ (units of
eV and \AA) was equally good for all states and we therefore used
these two parameters for all our calculations.

It may be surprising that such a simple inverse power potential can
reproduce the energetics of the liquid with sufficient accuracy, since
simple repulsive potentials cannot describe metallic bonding. One may
think that a more realistic potential such as those based on the
EAM~\cite{mei92,daw93,baskes92,belonoshko97} would be more
appropriate, since these potentials explicitly contain a repulsive and
a bonding term.  However, in our recent work on iron~\cite{alfe01a} we
tested the use of an EAM potential as a reference system and found
that the bonding term is almost independent of the positions of the
atoms, depending only on the volume and temperature of the system, and
the fluctuations of the energy are almost entirely due to the
repulsive term.  Since the only relevance in this work is the strength
of the fluctuations (Eq.~\ref{eqn:fluctuations}), little is gained by
using an EAM rather then a much simpler inverse power potential.

\subsection{Free energy of the reference system}

Consider the excess free energy of the IP, $F^{\rm ex}_{\rm IP} =
F_{\rm IP} - F_{\rm PG}$, where $F_{\rm PG}$ is the Helmholtz free
energy of the perfect gas and $F_{\rm IP}$ the total Helmholtz free
energy of the IP system.  The very simple functional form
of $U_{\rm IP}$ makes it easy to show that the adimensional quantity
$F^{\rm ex}_{\rm IP}/k_{\rm B}T$ can only depend non-trivially on a
single thermodynamic variable, rather then separately on $V$ and $T$:
\begin{equation}
F^{\rm ex}_{\rm IP}/k_{\rm B}T = f(\zeta)
\end{equation}  
with
\begin{equation}\label{eqn:zeta}
\zeta = B /V^{\alpha / 3} k_{\rm B} T.
\end{equation}  
The free energy of the IP has been studied extensively in the
past~\cite{laird92}, but only for special values of the exponent
$\alpha$, which did not include our own $\alpha=6.7$. We have
therefore explicitly calculated the free energy of our inverse power
potential using thermodynamic integration as before, but this time we
started from a system of known free energy, the Lennard-Jones liquid, whose
potential function is given by
\begin{equation}
U_{\rm LJ}=4\varepsilon \left[ \left( \frac \sigma r\right) ^{12}-\left( \frac
\sigma r\right) ^6\right].
\end{equation}
The free energy of the Lennard-Jones liquid, $F_{\rm LJ}$, has been
accurately tabulated by Johnson {\em et al.}~\cite{johnson93}.
To calculate $F_{\rm IP} - F_{\rm LJ} = \Delta F_{\rm LJ\rightarrow
IP}$ we used simulation cells containing 512 atoms with periodic
boundary conditions and a simulation time $T_{\rm sim}= 200$ ps.  We
performed the calculations for $\zeta$ ranging from 2.5 to 6.25, with
steps of 0.25. The calculations were done at a fixed volume of 14
\AA$^3$/atom and varing temperatures according to
Equation~\ref{eqn:zeta}.  We carefully checked that the results were
converged to better than 1 meV/atom with respect to the size of the
simulation cell and the length of the simulations.  To avoid
truncating the inverse power potential at a finite distance we used
the Ewald technique. Our results were fitted to a third order
polynomial in $\zeta$:
\begin{equation}
f(\zeta) = \sum_{i=0}^3 c_i \zeta^i.
\end{equation}
The coefficients are: $c_0 = 2.4333; c_1 = 27.805; c_2 = -5.0704; c_3
= 1.5177$, and the fitting function reproduced the calculated data
such that the errors in $F_{\rm IP}$ were generally less than 1 meV per
atom.


As an additional check on the calculated free energy, we repeated most
of the simulations using the perfect gas as the reference system,
thereby avoiding the inclusion of any possible errors that may exist
in the free energy of the LJ system reported in the
literature~\cite{johnson93}. For these calculations we used a
different form for $U_\lambda$, namely:
\begin{equation}
U_\lambda =  \lambda^2 U_{\rm IP} \; 
\end{equation}
(the potential energy of the perfect gas is zero, so does
not appear in the formula). So Eq.~\ref{eqn:ti} becomes
\begin{equation}
F_{\rm IP} - F_{\rm PG}  = \int_0^1 d \lambda \, 2\lambda \langle
U_{IP} \rangle_\lambda.
\end{equation}
The advantage of using this different functional form for $U_\lambda$
is that the value of the integrand does not need to be computed for
$\lambda=0$, where the dynamics of the system is determined by the
perfect gas potential. In this case, since there are no forces in the
system there is nothing stopping the atoms from overlapping, and the
potential energy $U_{\rm IP}$ diverges. Not computing the integrand at
$\lambda=0$ partially solves this problem, but for small values of
$\lambda$ where the forces on the atoms are small, the atoms can come
close together and the potential energy, $U_{\rm IP}$, fluctuates
violently. However, we found that by performing long enough
simulations, typically 1~ns, we could calculate the integral with an
accuracy of $\approx 1$ meV/atom, and, within the statistical
accuracy, we found the same results as those obtained using the LJ
reference system.

\subsection{Free energy of the {\em ab-initio} system}

To calculate the full {\em ab-initio} free energy of the liquid,
$F_{\rm liq}$, we used thermodynamic integration, starting from the IP
system. The calculations were performed at 18 different thermodynamic
states over a range of volumes (9.5-19.5 \AA$^3$/atom) and
temperatures (800-6000 K).  To address the issue of {\bf k}-point
sampling and cell size errors in the free energy difference $F_{\rm
liq} - F_{\rm IP}$, tests were carried out on cells containing up to
512 atoms and a $5 \times 5 \times 5$ {\bf k}-point grid (calculations
on the largest 512 atoms cell were only performed with
$\Gamma$-point sampling), at $V=16.5$ and $T=1000K$. The free energy
difference $F_{\rm liq} - F_{\rm IP}$ was calculated using the
perturbational approach (Eq.~\ref{eqn:secondorder}), with sets of
configurations generated using the IP potential. We found that a
64-atom cell with a $3\times 3 \times 3$ {\bf k}-point grid was
sufficient to get convergence to within 2 meV/atom. However, we were
reluctant to perform simulations using the desired $3\times 3 \times
3$ {\bf k}-point grid (14 points in the Brillouin Zone (BZ)) since
these calculations are extremely expensive.  We found it more
efficient to add one further step to our thermodynamic integration
scheme:
\begin{equation}\label{eqn:gammakappa}
\Delta F_{\Gamma \rightarrow 333}=F_{333}-F_\Gamma
=\int\limits_1^0d\lambda \left\langle U_{333}-U_\Gamma
\right\rangle_\lambda,
\end{equation}
where $U_{333}$ and $U_\Gamma$ are the {\em ab-initio} total energies
calculated using the $3\times 3 \times 3$ {\bf k}-point grid and
$\Gamma$-point sampling respectively, and $F_{333}$ and $F_{\Gamma}$
are the corresponding free energies. To evaluate the free energy
difference $\Delta F_{\Gamma \rightarrow 333}$ we noticed that the
difference $U_{333}-U_\Gamma$ did not depend significantly on the
position of the atoms, so the integral in Eq.~\ref{eqn:gammakappa}
could be evaluated using the second order formula
(Eq.~\ref{eqn:secondorder}). Using a long
$\Gamma$-point {\em ab-initio} simulation, we extracted up to 25 statistically
independent configurations and calculated the {\em ab-initio} energies
using the $3\times 3 \times 3$ {\bf k}-point grid. To test this, we performed spot checks at two thermodynamic
states, where we calculated the full thermodynamic integral $F_{333} -
F_{\rm IP} $ using adiabatic switching with a switching time of
$\approx 2$ ps, and found the same results to within a few meV/atom.

The free energy difference $\Delta F_{\rm IP \rightarrow \Gamma} =
F_\Gamma - F_{\rm IP}$ was obtained by full thermodynamic integration
between the {\em ab initio} and reference system using
adiabatic switching (Eq.~\ref{eqn:adiabatic_switching}) with a
switching time of 5 ps, which resulted in errors of 1(4) meV/atom in
the low(high) $P/T$ region.  To test this, we also calculated this free energy difference at several state points by numerical evaluation of the
thermodynamic integral (Eq.~\ref{eqn:ti}), with ~$\lambda$ = 0, 0.5 and 1; we found that this gave the same numerical answer to within our statistical errors.

In summary, the free energy of the liquid was obtained from a 
series of thermodynamic integration calculations:
\begin{equation}
F_{\rm liq} = F_{333} = F_{\rm LJ}+\Delta F_{\rm LJ\rightarrow
IP}+\Delta F_{\rm IP\rightarrow \Gamma }+\Delta F_{\Gamma \rightarrow
333}.
\end{equation}

\subsection{Representation of the free energy of the liquid}

The results of the calculations described in the previous section were
fitted to a suitable function of $T$ and $V$. In order to do that
efficiently we expressed the free energy in the following way:
\begin{equation}
F_{\rm liq} = F_{\rm IP} + \Delta F = F_{\rm IP} + \Delta U^s +
(\Delta F - \Delta U^s)
\end{equation}
where $\Delta U^s = U^s - U^s_{\rm IP}$, with $U^s $ the zero
temperature ab-initio (free) energy of the face-centred-crystal (fcc)
and $U^s_{\rm IP}$ the inverse power energy. $U^s$ can be calculated
very accurately, details of which will be given below in
Section~\ref{sec:perfect_freeenergy}; $U^s_{\rm IP}$ has no
errors. The remaining quantity $\Delta F - \Delta U^s$ is a
weak function of $V$ and $T$, and was fitted to a polynomial in $V$
and $T$:
\begin{equation}
\Delta F - \Delta U^s = \sum_{j=0}^1 \left ( \sum_{i=0}^{3} a_{ij} V^i
\right ) T^j
\end{equation}
The fitting reproduced the calculated data to within $\approx 2$
meV/atom.

\subsection{Error estimates for $F_{\rm liq}$}\label{sec:liquid_errors}
The errors on $F_{\rm IP}$ and $\Delta U^s$ are each less than 1
meV/atom (see Section~\ref{sec:perfect_freeenergy} below). The part of
the free energy that carries the largest errors is $\Delta F - \Delta
U^s$, which we estimate to be 2(5) meV/atom at
low(high) $P/T$.

\section{Free energy of the solid}\label{sec:solid}

The free energy of the solid can be represented as the sum of two
contributions: the free energy of the perfect
non-vibrating fcc crystal and that arising from atomic
vibrations above zero Kelvin:
\begin{equation}
F_{\rm sol} = F_{\rm perf} + F_{\rm vib}.
\end{equation}
The contribution to the free energy due to the vibrations of the atoms
may be written:
\begin{equation}
F_{\rm vib} = F_{\rm harm} + F_{\rm anharm}
\end{equation}
where $F_{\rm harm}$ is the free energy of the high temperature
crystal in the harmonic approximation and $F_{\rm anharm}$ is the
anharmonic contribution. 

\subsection{Free energy of the perfect crystal}\label{sec:perfect_freeenergy}

The free energy of the perfect crystal, $F_{\rm perf}$, was calculated
as a function of volume and temperature. Calculations were performed on a 
fcc cell at a series of volumes (9.5-19.5~\AA$^3$/atom representing
compression up to $\sim 150$ GPa) and temperatures (up to 6000K) with a
24x24x24 {\bf k}-point grid (equivalent to 1300 points in the
irreducible wedge of the Brilloiun zone (IBZ)), which ensures
convergence of the (free) energies to better than 1 meV/atom. At each
different temperature we calculated the {\em ab-initio} (free)
energy as a function of volume, and then performed a least-square fit of the
results to a third-order Birch-Murnaghan equation of state:
\begin{eqnarray}\label{murna}
E(V) = E_0 + \frac{3}{2}V_0K \left [ \frac{3}{4}(1+2\xi)\left
(\frac{V_0}{V}\right )^{4/3} - \frac{\xi}{2}
\left ( \frac{V_0}{V} \right )^{2} 
 -\frac{3}{2}(1+\xi) \left ( \frac{V_0}{V}
\right )^{2/3} +
\frac{1}{2} \left ( \xi + \frac{3}{2}\right ) \right ] \\
\xi = \frac{3}{4}(4 - K'). \hspace{12cm} \nonumber
\end{eqnarray}
The parameters $E_0, V_0, K_0,$ and $K'$ were fitted to a
fourth order polynomial as function of temperature:
\begin{equation}
E_0(T) = \sum_{i=0}^{4} e_{0,i} T^i; \hskip 20pt
V_0(T) = \sum_{i=0}^{4} v_{0,i} T^i; \hskip 20pt
K_0(T) = \sum_{i=0}^{4} k_{0,i} T^i; \hskip 20pt
K'(T) = \sum_{i=0}^{4} k_{0,i}' T^i. 
\end{equation}
The fitting reproduced the calculated energies to better than 1
meV/atom in the whole $P/T$ range.

\subsection{Free energy of the harmonic crystal}

The free energy of the harmonic crystal is given by:
\begin{equation}
F_{\rm harm}(V,T)=-\left( \frac{3k_{\rm B}T}{\Omega _{\rm BZ}N_i}\right)
\sum\limits_i\int\limits_{\rm BZ}\left( \ln \left[ \frac{k_BT}{\hbar
\omega _{{\bf q},i}(V,T)}\right] -\frac {1}{24}\left[ \frac{\hbar \omega_{{\bf
q},i}(V,T)}{k_{\rm B}T}\right] ^2+\dots \right) d{\bf q}
\end{equation}
where $\omega _{{\bf q},i}(V,T)$ are the phonon frequencies of branch
{\it i} and wavevector {\bf q}, $\Omega _{\rm BZ}$ is the volume of
the Brillouin zone, $N_i$ is the total number of phonon branches and
the dependence on temperature of $\omega _{{\bf q},i}$ is due to
electronic excitations.  We truncate the summation after the first term, which
is the classical limit of the free energy:
\begin{equation}
F_{\rm harm}=-\left( \frac{3k_{\rm B}T}{\Omega _{\rm BZ}N_i}\right)
\sum\limits_i\int\limits_{\rm BZ}\left( \ln \frac{k_{\rm B}T}{\hbar \omega
_{{\bf q},i}} \right) d{\bf q}.
\end{equation}
This is a justifiable approximation to make for two reasons: (i) the
error in making such a truncation is very small ($<$1 meV/atom), and
(ii) neglecting the higher order terms, i.e., the quantum corrections,
is consistent with the liquid calculations where the motions of the
atoms were treated classically.

It is useful to express the harmonic free energy in terms of the
geometric average $\bar{\omega}$ of the phonon frequencies, defined
as:
\begin{equation}
\ln \bar{\omega} = \frac{1}{N_{{\bf q}} N_i} \sum_{{\bf q}, i}
\ln ( \omega_{{\bf q} i} ) ,
\end{equation}
where we have replaced the integral $\frac{1}{\Omega_{\rm
BZ}}\int\limits_{\rm BZ} d{\bf q}$ with the summation $\frac{1}{N_{\bf
q}}\sum\limits_{\bf q}$.
This allows us to write:
\begin{equation}
F_{\rm harm} = 3 k_{\rm B} T \ln ( \beta \hbar \bar{\omega} ).
\end{equation}

To calculate the vibrational frequencies $\omega_{{\bf q},i}$, we used
our own implementation~\cite{darioweb} of the small displacement
method~\cite{kresse95,alfe01a}. 

The central quantity in the calculation of the phonon frequencies
is the force-constant matrix $\Phi_{i s \alpha , j t \beta}$,
since the frequencies at wavevector ${\bf q}$ are
the eigenvalues of the dynamical matrix $D_{s \alpha , t \beta}$,
defined as:
\begin{equation}
D_{s \alpha , t \beta} ( {\bf q} ) = \frac{1}{\sqrt{M_s M_t}} \sum_i
\Phi_{i s \alpha , j t \beta} \exp \left[ i {\bf q} \cdot ( {\bf
R}_j^0 + {\bf \tau}_t - {\bf R}_i^0 - {\bf \tau}_s) \right] \; .
\end{equation}
where ${\bf R}_i^0$ is a vector of the lattice connecting different
primitive cells, ${\bf \tau}_s$ is the position of the atom $s$ in the
primitive cell and $M_s$ its mass. If we have the complete
force-constant matrix, then $D_{s \alpha , t \beta}$, and hence the
frequencies $\omega_{{\bf \rm q} l}$, can be obtained at any ${\bf q}$.  In
principle, the elements of $\Phi_{i s \alpha , j t \beta}$ are
non-zero for arbitrarily large separations $\mid {\bf R}_j^0 + {\bf
\tau}_t - {\bf R}_i^0 - {\bf \tau}_s \mid$, but in practice they decay
rapidly with separation, so a key issue in achieving our target
precision is the cut-off distance beyond which the elements can be
neglected.

In the harmonic approximation the $\alpha$ Cartesian component of
the force exerted on the atom at position ${\bf R}_i^0 + {\bf \tau}_s$
is given by:
\begin{equation}
F_{i s \alpha} = - \sum_{j t \beta} \Phi_{i s \alpha , j t \beta}~u_{j t \beta}
\end{equation}
where $u_{j s \beta}$ is the displacement of the atom in ${\bf R}_j^0
+ {\bf \tau}_t$ along the direction $\beta$.  The force constant matrix can
be calculated {\em via}:
\begin{equation}\label{displ}
\Phi_{i s \alpha, j t \beta}=-\frac{F_{i s \alpha, j t \beta}}{ u_{j t \beta} }
\end{equation}
where all the atoms of the lattice are displaced one at a time along the
three Cartesian components by $u_{j t \beta}$, and the
forces $F_{i s \alpha, j t \beta}$ induced on the atoms in ${\bf
R}_i^0 + {\bf \tau}_s$ are calculated.  Since the crystal is invariant under
translations of any lattice vector, it is only necessary to displace
the atoms in one primitive cell and calculate the forces induced on
all the other atoms of the crystal, so that we can simply put $j = 0$.
The fcc crystal has only one atom in the
primitive cell, so only three displacements are needed. However, a
displacement along the $x$ direction is equivalent by symmetry to a
displacement along the $y$ or the $z$ direction, and therefore only
one displacement along an arbitrary direction is needed. It is
convenient to displace the atom along a direction of high symmetry, so
that the supercell has the maximum possible number of symmetry
operations. These can be used to reduce the number of {\bf k}-points
in the IBZ, minimising the computational effort. For an
fcc crystal this is achieved by displacing the atom along the
diagonal of the cube.

Tests for cell-size (64-512 atoms), displacement length (0.01-0.0005
fraction of nearest neighbours distance) and {\bf k}-point grid (up to
$9 \times 9\times 9$) were performed at the two extremes of high $P/T$
and low $P/T$ state points. Convergence of the free energy to within less
than 3 meV/atom was achieved using a 64-atom cell with a 0.001
fractional displacement and a $9 \times 9\times 9$ k-point grid
(equivalent to 85 points in the IBZ of the supercell). Calculations
were performed for $V= 9.5-18.5~{\rm \AA}^3$ and $T=500-6000K$, and
$\ln (\bar{\omega})$ has been fitted to the following polynomial in
$V$ and $T$:
\begin{equation}
\ln (\bar{\omega}) = \sum_{j=0}^3 \left ( \sum_{i=0}^{3} a_{ij} V^i \right ) T^j.
\end{equation}
The fitting reproduced the calculated data within $\approx 1$
meV/atom.

\subsection{Anharmonicity}

To obtain the anharmonic contribution to the free energy of the solid
we have again used thermodynamic integration. In this case a natural
choice for the reference system could be the harmonic
solid~\cite{dewijs98}, but unfortunately this does not reproduce the
{\it ab initio} anharmonic system with sufficient accuracy. A much
better reference system is a linear combination of the harmonic {\em
ab-initio} and the same IP used for the liquid calculations~\cite{alfe01a}:
\begin{equation}
U_{\rm ref} = a U_{\rm IP} + b U_{\rm harm},
\end{equation}
where the harmonic potential energy is:
\begin{equation}
U_{\rm harm} = \frac 1 2 \sum_{i s \alpha, j t \beta}~u_{i s \alpha}
\Phi_{i s \alpha , j t \beta}~u_{j t \beta},
\end{equation}
and where $u_{j s \beta}$ is the displacement of the atom in ${\bf R}_j^0 +
{\bf \tau}_t$ along the direction $\beta$, and $\Phi_{i s \alpha , j t
\beta}$ is the force constant matrix.  The parameters $a$ and $b$ are
determined by minimising the fluctuations in the energy differences
$U_{\rm ref} - U$ on a set of statistically independent configurations
generated with $U_{\rm ref}$. However, when we start our optimisation
procedure we do not know $U_{\rm ref}$, so we cannot use it to
generate the configurations. We could use the {\em ab-initio}
potential, but this would involve very expensive calculations.  We
used instead an iterative procedure, like in our previous work on
iron~\cite{alfe01a}. We generated a set of configurations using the
harmonic potential $U_{\rm harm}$ and calculated the {\em ab-initio}
energies. By minimising the fluctuations of $U_{\rm ref} - U$ we found
a first estimate for $a$ and $b$, and we constructed a first estimate
of $U_{\rm ref}$. We generated a second set of configurations using
this $U_{\rm ref}$, calculated the {\em ab-initio} energies and
minimised again the fluctuations of $U_{\rm ref} - U$ with respect to
$a$ and $b$. This procedure could be continued until the values of
$a$ and $b$ no longer changed, but in practice we stopped after the
second step, and found $a=0.95$ and $b=0.12$.  We did not use the
extra freedom in the choice of the inverse power parameters since we
found that this reference system already described the energetics of
the solid very accurately.

The calculation of the anharmonic part of the free energy required,
once more, two thermodynamic integration steps. In the first step we
calculated the free energy difference $F_{\rm ref} - F_{\rm
harm}$. These are cheap calculations since they involve only the classical
potentials $U_{\rm IP}$ and $U_{\rm harm}$; the simulations were
performed with cells containing 512 atoms for 10 ps, which ensured
convergence of the free energy difference $F_{\rm ref} - F_{\rm harm}$
to within 1 meV/atom.  In the second step we calculated $F_{\rm vib} -
F_{\rm ref}$ where, since the fluctuations in the energy differences $U - U_{\rm ref}$
were very
small, we were able to use the second order formula
(Eq.~\ref{eqn:secondorder}).


The problem in the calculation of thermal averages for a nearly
harmonic system is that of ergodicity. For an harmonic system
different degrees of freedom do not exchange energy, so in a system
which is close to being harmonic the exploration of phase space using
molecular dynamics can be a very slow process. We solved this problem
following Ref.~\cite{dewijs98} whereby the statistical sampling was
performed using Andersen molecular dynamics~\cite{andersen80}, in
which the atomic velocities are periodically randomised by drawing
them from a Maxwellian distribution. This type of simulation generates
the canonical ensemble and overcomes the ergodicity problem.

All the calculations were performed on a 64-atom cell with kpoints in
a $7\times 7 \times 7$ grid for the high P/T state points and a
$9\times 9 \times 9$ grid for the low P/T state points equivalent to
172 or 365 points in the IBZ respectively.

The anharmonic contribution to the free energy of the solid turns out
to be very small, being positive and equal to only a few meV/atom at
low pressure and approximatively -20 meV/atom at high pressure.

\subsection{Error estimates for $F_{\rm sol}$}
The errors in $F_{\rm perf}$ are less than 1 meV/atom, the errors in
$F_{\rm harm}$ are $\approx 3(4)$ meV/atom at low(high) $P/T$ and the
errors in $F_{\rm ahnarm}$ are $\approx 1(4)$ meV/atom at low(high)
$P/T$; the total errors in $F_{\rm sol}$ are $\approx 3(6)$ meV/atom
at low(high) $P/T$.

\section{Results and discussion}\label{sec:discussion}

We display in Figure 1 our calculated melting curve compared with the
experimental zero pressure value~\cite{crc97}, the DAC high pressure
results~\cite{boehler97, hanstrom00} and the high pressure shock
datum~\cite{shaner84}. We also report in Figures 2a, 2b and 2c the
volume change on melting, $V_m$, the entropy change on melting, $S_m$,
and the melting gradient, $dT_m/dP$, respectively. The errors in the
melting curve arise from the errors in the calculated free energies
and are $\approx 50(100) $ K in the low(high) pressure part of the
diagram respectively.

The overall agreement with the experiments is extremely good; however,
the low pressure results differ by more than 15 \% (at zero pressure,
786 K compared with the experimental value of 933 K). Indeed, at zero
pressure the agreement between the calculated and experimental volume
change on melting and $dT_m/dP$ is rather poor (see
Table~\ref{tab:meltprops}). In addition, our calculations are not in
very good agreement with the previous calculations of de Wijs {\em et
al.}~\cite{dewijs98}, although this is not necessarily surprising,
since these latter calculations were based on LDA, while ours are
based on GGA. Nevertheless, one might expect the results from LDA and
GGA to be similar, since Al is a nearly free-electron like metal and
therefore one would expect a very good DFT description with both LDA
and GGA.
To explore a possible reason why GGA does not predict the melting
properties of aluminium very accurately we consider the zero pressure
crystal equilibrium volume. This is predicted by GGA to be $\approx 2
\%$ larger than the experimental value; this means that the calculated
pressure for the experimental zero pressure volume is $\approx +1.6$
GPa.  
To see how this error propagates in melting properties we may
devise a correction to the Helmholtz free energy such that the
pressure is rectified:
\begin{equation}
F_{\rm corr} = F + \delta P V,
\end{equation}
with $\delta P = 1.6$ GPa. Using $F_{\rm corr}$ in our calculations we
found the {\em corrected} melting curve, represented by the dotted
line in Fig. 1, where we assumed $\delta P$ to be the same in the
whole $P/T$ range. The zero pressure corrected melting temperature is
912 K, which is in very good agreement with the experimental value 933
K. The corrected volume change on melting, entropy change on melting
and $dT_m/dP$ are also in much better agreement with the experimental
numbers. The correction is less important at high pressure, where
$dT_m/dP$ is smaller.

This point may be further illustrated by looking at the zero pressure
phonon dispersion curves for Al. Since phonon frequencies depend on
the interatomic forces, their correctness is surely important in the
context of melting.  In Figure 3 we display the GGA calculated phonon
dispersion curves compared with experimental
data~\cite{stedman66}. Our calculations were performed both at the GGA
zero pressure equilibrium volume and the experimental volume (both
at 80 K).  We notice that the agreement is good (though not perfect)
if the calculations are performed at the experimental volume, and
rather poor if the calculated zero pressure GGA volume is used
instead.  This indicates that GGA will probably yield better results
if the GGA pressure is corrected in order to match the experimental
data.

In their work, de Wijs {\em et al.}~\cite{dewijs98} found good
agreement between LDA and experiments. In their case a {\em corrected}
LDA would lower the zero pressure melting point below 800 K. In order
to understand this apparent different behaviour between LDA and GGA we
have also calculated the phonons using LDA at the calculated
equilibrium volume and also at the experimental volume (both at
80K). These are also reported in Fig. 3. In accord with previous LDA
calculations~\cite{degironcoli95} we found very good agreement with
the experiments when the phonons are calculated at the LDA zero
pressure volume, but the agreement becomes poor at the experimental
volume, which is consistent with the result for the melting
temperature~\cite{dewijs98}.

In conclusion, both GGA and LDA predict an incorrect equilibrium volume
at a fixed pressure, although LDA yields very good results for both
the phonon dispersion curves and the zero pressure melting properties
(which is probably accidental). For GGA the incorrect equilibrium volume
propagates to an incorrect description of the phonon frequencies and
the melting properties. If the GGA pressures are corrected so as to
match the experimental data, the phonon dispersion and the melting
properties come out in very good agreement with the experiments. These
two behaviours are internally consistent, but point to an intrinsic
error due to the use of GGA.  Quantum Monte-Carlo (QMC)
techniques~\cite{hammond94} have been shown to predict the energetics
with much higher accuracy than DFT~\cite{rajagopal95}, and
calculations for systems containing more than 100 atoms have already
been reported~\cite{kent99}. We believe that in the near future it
will be possible to use QMC for more accurate calculations of free
energies.

To summarise, we have calculated the melting curve of aluminium
entirely from first-principles within the DFT-GGA framework. Our work
is based on the calculation of the Gibbs free energy of liquid and
solid Al, and for each fixed pressure the melting temperature is
determined by the point at which the two free energies cross. Our
results are in good agreement with the available experimental data,
although they reveal an intrinsic DFT-GGA error which is responsible
for an error of $\approx 150 $K in the low pressure melting
curve. This error is probably due to the incorrectly predicted
pressure by GGA, and it becomes less important in the high
pressure region, as $dT_m/dP$ becomes smaller.




\section*{Acknowledgements} 

We both acknowledge the support of Royal Society University Research
Fellowships; we also thank Mike Gillan and John Brodholt for useful discussions. LV thanks Humphrey Vocadlo for his assistance during the course of this research.

\pagebreak
\newpage

\begin{table}
\begin{tabular}{l|ccccc}
  &  Experiment  & LDA  & GGA & GGA - corrected \\
\hline 
$T_m$ (K)& 933 & 890 (20) & 786 (50) & 912 (50) \\
$S_m$ ($k_{\rm B}$)& 1.38 & 1.36 (4) & 1.35 (6) & 1.37 (6) \\
$V_m$ (\AA$^3$)& 1.24 & 1.26 (20) & 1.51 (10) & 1.35 (10) \\
$dT_m/dP$ (K~GPa$^{-1}$) & 65 & 67 (12) & 81 & 71 \\
\end{tabular}
\caption{Comparison of {\em ab initio} and experimental melting
properties of Al at zero pressure. Values are given for the melting
temperature, $T_m$, entropy change on melting, $S_m$, volume change on
melting, $V_m$, and melting gradient $dT_m/dP$. The LDA results are
from Ref.~\protect\cite{dewijs98}; the experimental values for $T_m$, $S_m$ and $dT_m/dP$ are from Refs. ~\cite{crc97}, ~\cite{chase85} and ~\cite{cannon74} respectively, and the experimental melting volume, $V_m$, is calculated using the Clapeyron relation, $V_m = S_mdT_m/dP$.} 
\label{tab:meltprops}
\end{table}

\pagebreak
\newpage

\begin{figure}
\psfig{figure=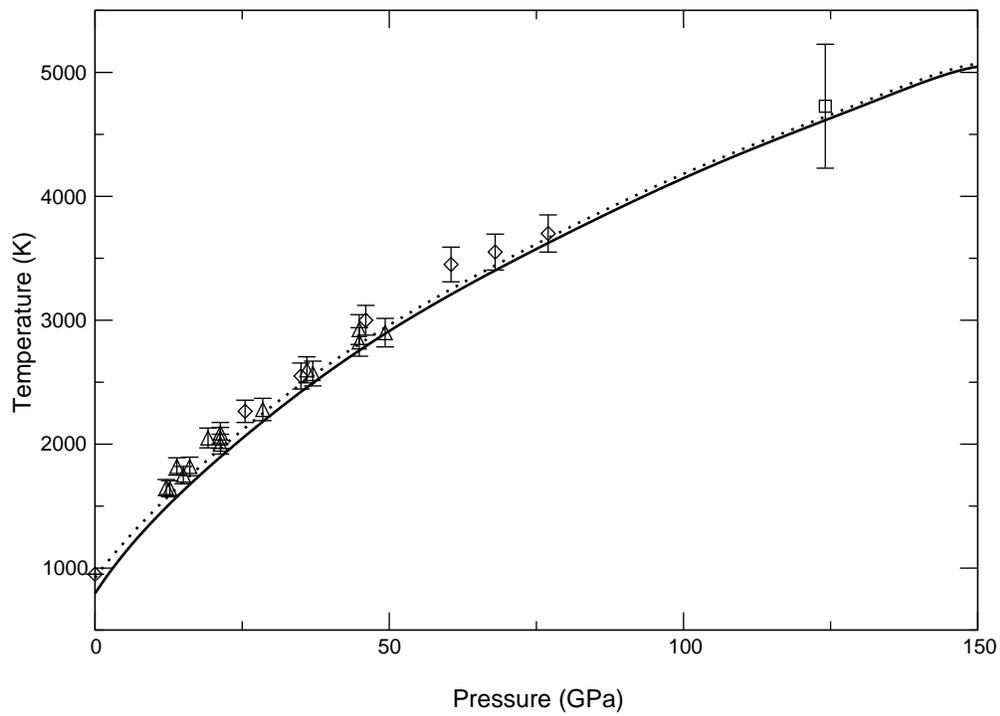,height=6in}
\caption{Comparison of melting curve of Al from present calculations
with previous experimental results. Solid curve: present work;  dotted
curve: present work with pressure correction (see text); diamonds and
triangles: DAC measurements of Refs.~\protect\cite{boehler97} and
~\protect\cite{hanstrom00} respectively; square: shock experiments of
Ref.~\cite{shaner84}.}
\label{fig:melting_curve}
\end{figure}

\begin{figure}
\psfig{figure=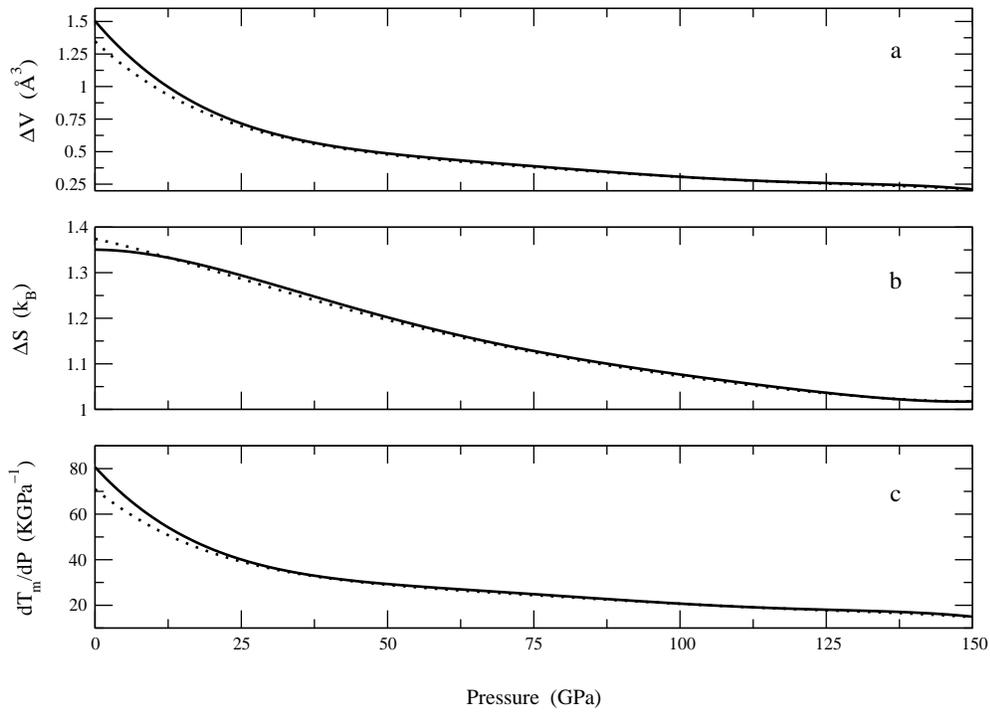,height=6in}
\caption{Calculated pressure dependence of the melting properties of
Al: a) volume change on melting, b) entropy change on melting and c)
melting gradient.  Solid curve: present work; dotted curve: present
work with pressure correction (see text).}
\label{fig:melting_props}
\end{figure}

\begin{figure}
\psfig{figure=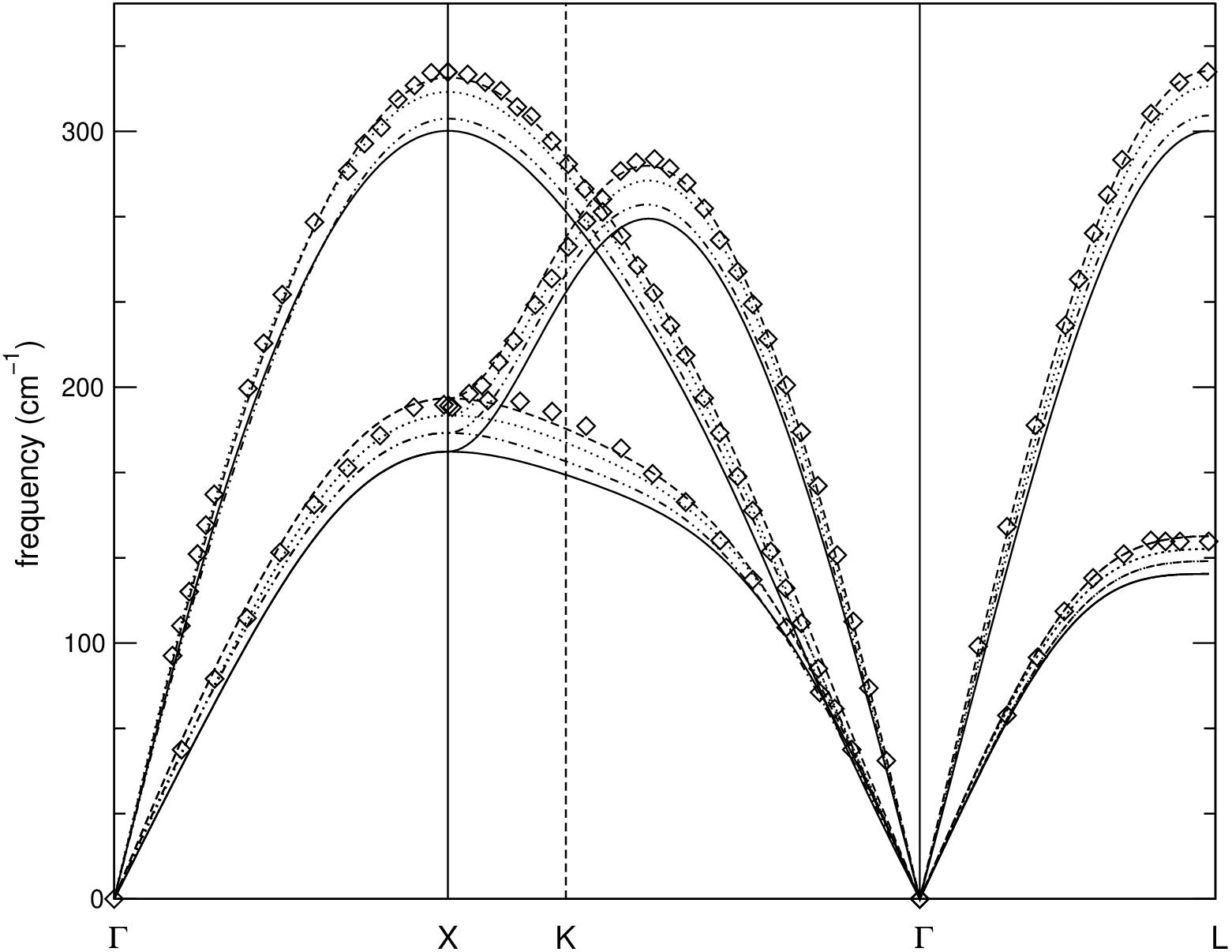,height=6in}
\caption{Comparison of the phonon dispersion curve for Al from present
calculations with previous experimental results. Solid curves: present
work with GGA; dotted curves: present work with GGA and with pressure
correction (see text); dashed curves: present work with LDA;
dot-dashed curves: present work with LDA and with pressure correction
(see text); diamonds: experiments from Ref.~\cite{stedman66}.}
\label{fig:phonon_curve}
\end{figure}


\begin{thebibliography}{99}

\bibitem{williams87} Q. Williams, R. Jeanloz, J. D. Bass,
B. Svendesen, T. J. Ahrens, Science {\bf 286}, 181 (1987).

\bibitem{boehler93} R. Boehler, Nature {\bf 363}, 534 (1993).

\bibitem{shen98} G. Shen, H. Mao, R. J. Hemley, T. S. Duffy and
M. L. Rivers, Geophys. Res. Lett. {\bf 25}, 373 (1998).



\bibitem{brown86} J. M. Brown and R. G. McQueen, J. Geophys. Res. {\bf
91}, 7485 (1986).

\bibitem{yoo93} C. S. Yoo, N. C. Holmes, M. Ross, D. J. Webb and
C. Pike, Phys. Rev. Lett. {\bf 70}, 3931 (1993).

\bibitem{alfe99} D. Alf\`e, M. J. Gillan and G. D. Price, Nature {\bf
401}, 462 (1999).

\bibitem{alfe01a} D. Alf\`e, G. D. Price, M. J. Gillan, Phys. Rev. B,
in press.

\bibitem{alfe01b} D. Alf\`e, G. D. Price, M. J. Gillan, unpublished.

\bibitem{laio00} A. Laio, S. Bernard, G. L. Chiarotti, S. Scandolo and
E. Tosatti, Science {\bf 287}, 1027 (2000).

\bibitem{belonoshko00} A. B. Belonoshko, R. Ahuja, and B. Johansson,
Phys. Rev. Lett. {\bf 84}, 3638 (2000).

\bibitem{crc97} {\it CRC Handbook of Chemistry and Physics}, Editor
D. R. Lide, New York, 77th edition (1996-1997).

\bibitem{boehler97} R. Boehler, M. Ross, Earth Planet. Sci. Lett. {\bf
153}, 223 (1997)

\bibitem{hanstrom00} A. H\"{a}nstr\"{o}m, P. Lazor, J. of Alloys and
Compounds {\bf 305}, 209 (2000).

\bibitem{shaner84} J. W. Shaner, J. M. Brown, R. G. McQueen, in {\rm
High pressure in Science and Technology}. Editors C. Homan, R. K. Mac
Crone, E. Whalley, North Holland, Amsterdam, 137 (1984).

\bibitem{moriarty84} J. A. Moriarty, D. A. Young, and M. Ross,
Phys. Rev. B {\bf 30}, 578 (1984).


\bibitem{mei92} J. Mei, J. W. Davenport, Phys. Rev. B {\bf 46}, 21
(1992).

\bibitem{morris94} J. R. Morris, C. Z. Wang, K. M. Ho, and C. T. Chan,
Phys. Rev. B {\bf 49}, 3109 (1994).

\bibitem{straub94} G. K. Straub, J. B. Aidun, J. M. Willis,
C. R. Sanchez-Castro, and D. C. Wallace, Phys. Rev. B {\bf 50}, 5055
(1994).

\bibitem{dewijs98} G. A. de Wijs, G. Kresse and M. J. Gillan,
Phys. Rev. B {\bf 57}, 8223 (1998).

\bibitem{generaldft} P. Hohenberg and W. Kohn, Phys. Rev. {\bf 136},
B864 (1964); W. Kohn and L. Sham, Phys. Rev. {\bf 140}, A1133 (1965);
R. O. Jones and O. Gunnarsson, Rev. Mod. Phys. {\bf 61}, 689 (1989);
M. J. Gillan, Contemp. Phys. {\bf 38}, 115 (1997).

\bibitem{frenkel96} For a discussion of thermodynamic integration see
e.g.  D. Frenkel and B. Smit, {\em Understanding Molecular
Simulation}, Academic Press, San Diego (1996).

\bibitem{jesson2000} B. J. Jesson and P. A. Madden, J. of
Chem. Phy. {\bf 113} 5924 (2000).

\bibitem{wang91} Y. Wang and J. Perdew, Phys. Rev. B {\bf 44}, 13298
(1991).

\bibitem{perdew92} J. P. Perdew, J. A. Chevary, S. H. Vosko,
K. A. Jackson, M. R. Pederson, D. J. Singh and C. Fiolhais,
Phys. Rev. B {\bf 46}, 6671 (1992).

\bibitem{mermin65} N. D. Mermin, Phys. Rev. {\bf 137}, A1441 (1965).

\bibitem{gillan89} M. J. Gillan, J. Phys. Condens. Matter {\bf 1}, 689
(1989).

\bibitem{wentzcovitch92} R. M. Wentzcovitch, J. L. Martins and
P. B. Allen, Phys. Rev. B {\bf 45}, 11372 (1992).

\bibitem{kresse96a} G. Kresse and J. Furthm\"{u}ller, Phys. Rev. B
{\bf 54}, 11169 (1996); a discussion of the ultra-soft
pseudopotentials used in the VASP code is given in G. Kresse and
J. Hafner, J. Phys. Condens. Matter {\bf 6}, 8245 (1994).

\bibitem{vanderbilt90} D. Vanderbilt, Phys. Rev. B {\bf 41}, 7892
(1990).

\bibitem{monkhorst76} H. J. Monkhorst and J. D. Pack, Phys. Rev. B
{\bf 13}, 5188 (1976).

\bibitem{alfe99b} D. Alf\`{e}, Comp. Phys. Commun. {\bf 118}, 31
(1999).

\bibitem{sugino95} O. Sugino and R. Car, Phys. Rev. Lett. {\bf 74},
1823 (1995).

\bibitem{smargiassi95a} E. Smargiassi, P. A. Madden, Phys Rev B {\bf
51}, 117 (1995).

\bibitem{alfe00} D. Alf\`{e}, G. A. de Wijs, G. Kresse and
M. J. Gillan, Int. J. Quant. Chem. {\bf 77}, 871 (2000).

\bibitem{nose84} S. Nos\'e, Molec. Phys., {\bf 52} 255 (1984);
J. Chem. Phys. {\bf 81}, 511 (1984).

\bibitem{ditolla93} F. D. Di Tolla and M. Ronchetti, Phys. Rev. B {\bf
48}, 1726 (1993).

\bibitem{watanabe90} M. Watanabe and W. P. Reinhardt,
Phys. Rev. Lett. {\bf 65}, 3301 (1990).

\bibitem{daw93} M. S. Daw, S. M. Foiles, and M. I. Baskes,
Mat. Sci. Rep. {\bf 9}, 251 (1993).

\bibitem{baskes92} M. I. Baskes, Phys. Rev. B {\bf 46}, 2727 (1992).

\bibitem{belonoshko97} A. B. Belonoshko, and R. Ahuja, Phys. Earth
Planet. Inter. {\bf 102}, 171 (1997).

\bibitem{laird92} B. B. Laird and A. D. J. Haymet, Mol. Phys. {\bf
75}, 71 (1992).
 
\bibitem{johnson93} K. Johnson, J. A. Zollweg, and E. Gubbins,
Mol. Phy. {\bf 78}, 591 (1993).

\bibitem{darioweb} Program available at {\tt
http://chianti.geol.ucl.ac.uk/\~dario}

\bibitem{kresse95} G. Kresse, J. Furthm\"{u}ller and J. Hafner,
Europhys. Lett. {\bf 32} 729 (1995).

\bibitem{andersen80} H. C. Andersen, J. Chem. Phys. {\bf 72}, 2384
(1980).

\bibitem{stedman66} R. Stedman, and G. Nilsson, Phys. Rev. {\bf 145},
492 (1966).



\bibitem{degironcoli95} S. de Gironcoli, Phys. Rev. B. {\bf 51}, 6773
(1995)

\bibitem{hammond94} B. L. Hammond, W. A. Lester, Jr., \&
P. J. Reynolds, {\em Monte Carlo Methods in Ab Initio Quantum
Chemistry}, World Scientific, Singapore, (1994).

\bibitem{rajagopal95} G. Rajagopal, R. J. Needs, A. James,
S. D. Kenny, \& W. M. C. Foulkes, Phys. Rev. B {\bf 51}, 10591 (1995).

\bibitem{kent99} P. R. C. Kent, R. Q. Hood, A. J. Williamson,
R. J. Needs, W. M. C Foulkes, G. \& Rajagopal, Phys. Rev. B {\bf 59},
1917 (1999).

\bibitem{cannon74} J. F. Cannon, J. Phys. Chem. Ref. Data 3, 781 (1974).
\bibitem{chase85} M. W. Chase, Jr., C. A. Davies, J. R. Downey, Jr., D. J. Frurip, R. A. McDonald and A. N. Syverud, J. Phys. Chem. Ref. Data Suppl. 14, 1 (1985).
\end{thebibliography}
\end{document}